\documentclass[prl,aps,twocolumn,amssymb,amsmath]{revtex4}
\usepackage{graphicx}
\newcommand{\VEC}[1]{{\mathbf{#1}}}
\newcommand{\x}{\VEC{x}}

\newcommand{\rmd}{\mathrm{d}}
\newcommand{\rme}{\mathrm{e}}

\newcommand{\z}{\ensuremath{\zeta}}
\newcommand{\g}{\ensuremath{\text{gauss}}}
\newcommand{\mean}[1]{\langle #1 \rangle}
\newcommand{\sample}[1]{\overline{#1}}
        \newcommand{\fig}[2]{\includegraphics[width=#1]{#2.eps}}
        \newcommand{\Fig}[1]{\includegraphics[width=\columnwidth]{#1.eps}}
\usepackage{times}

\begin{document}
\title{
Universal interface width distributions at the depinning threshold
}
\author{Alberto Rosso$^{1}$, Werner Krauth$^{1}$, Pierre Le
Doussal$^{2}$, Jean
Vannimenus$^{1}$ and Kay J\"org Wiese$^{3}$}
\affiliation{$^{1}$CNRS-Laboratoire de Physique Statistique de
l'Ecole Normale Sup{\'e}rieure, 24 rue Lhomond, 75231 Paris, France\\
$^{2}$CNRS-Laboratoire de Physique Th{\'e}orique de
l'Ecole Normale Sup{\'e}rieure, 24 rue Lhomond, 75231 Paris, France\\
$^{3}$KITP, University of California at
Santa Barbara, Santa Barbara, CA 93106, USA}     

\begin{abstract}
We compute the probability distribution of the interface width
at the depinning threshold, using recent powerful algorithms. It
confirms the universality classes found previously.  In all cases,
the distribution is surprisingly well approximated by a generalized
Gaussian theory of independant modes which decay with a characteristic
propagator $G(q)=1/q^{d+2\z}$; $\z$, the roughness exponent, is computed
independently. A functional renormalization analysis explains this
result and allows to  compute the small deviations, i.e.  a universal
kurtosis ratio, in agreement with numerics. We stress the importance of
the Gaussian theory to interpret numerical data and experiments.
\pacs{64.60.Lx, 05.40.+j, 05.70.Ln}
\end{abstract}
\maketitle

The scaling properties of driven elastic interfaces in random media
play an important role in a wide variety of physical situations,
ranging from stochastic surface growth to domain walls in disordered
magnetic materials, the spreading of fluids on rough substrates, and
the dynamics of cracks
\cite{Stanley.Fractal_concepts.95,Kardar.dynamics.98}.  These problems
share many features with critical phenomena and provide a challenge
for theoretical approaches to disordered systems and non-equilibrium
phenomena
\cite{Nattermann.pinning.97,Fisher.Collective.98,Narayan.Pinning.93,LeDoussalWiese2002a,Wiese.2loop.01,Vannimenus.interface.01}.

\begin{figure}[t]
\centerline{\fig{0.7\columnwidth}{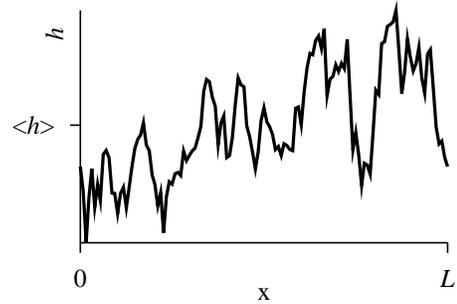} } 
\caption{Example of a $1+1$-dimensional
periodic interface $h(x)$ (random walk) with mean value
$\langle h \rangle$, and $u(x) = h(x) - \langle h \rangle$.} 
\label{f:schema}
\end{figure}

Here we study interfaces described by a scalar height function
$h(\x)$, where $\x$ is the $d$-dimensional internal coordinate.  We
measure the deviation from the mean position as $u(\x) =
h(\x) - \mean{h}$, where $\mean{\ldots}$ stands for the spatial
average over all $\x$ of a given interface (cf.~Fig.~\ref{f:schema}).
The mean square width of a \emph{single} interface, $w^2(\{u(\x)\}) = \mean{u^2}$,
can be used to characterize its roughness, and explore universal
properties: After averaging over the ensemble of interfaces, 
$w^2$ grows with the lateral extension $L$ of the system as
\begin{equation}
 \sample{w^2} \propto L^{2 \zeta}\quad \text{for}\quad L \rightarrow
\infty\ , \label{e:zeta}
\end{equation}
where $\z$ is the roughness exponent.

An interesting property is that, for positive $\z$, $w^2$ fluctuates
even in the thermodynamic limit
\cite{Racz.Random.94,Racz.Growth.94,Plischke.Curvature.94}. This means
that the long-range geometric features of the interface are not characterized by
the roughness exponent alone, but require the complete
probability distribution $P(w^2)$. $P(w^2)$ has been
computed for several linear stochastic growth equations without
disorder as the Edwards-Wilkinson model, the Mullins-Herring model,
and the $1$-d KPZ model
\cite{Plischke.Curvature.94,Racz.Growth.94}. In these models, the
probability distribution
$P(w^2)$ can be rescaled into a form independent of 
system size and of microscopic details
\begin{equation}
P(w^2) = (1/\sample{w^2}) \Phi(w^2/\sample{w^2}) \quad \text{for} 
\quad L \rightarrow \infty .
\end{equation}
Although $\sample{w^2}$ may contain a non-universal scale, the
function $\Phi(z)$ is universal.  It has been argued that the shape of
$P(w^{2})$ can thus be used as a sensitive tool, distinct from
$\zeta$, to distinguish between different universality classes
\cite{Racz.Random.94,Racz.Growth.94,Plischke.Curvature.94,Antal.1overf.01}.
Furthermore, $\Phi(z)$ is expected to converge to a $\delta$-function
above the
upper critical dimension $d_{\mathrm{\text{uc}}}$. This has motivated
attempts to determine  $d_{\mathrm{\text{uc}}}$ for e.g.\ the KPZ equation
\cite{Marinari.KPZ.02}. Probability
distributions of order parameters have  received much attention for
related models such as polymers, spin glasses, and random diffusion
\cite{bouchaud.georges}.  The quantity we study here, $P(w^{2})$, is
the distribution of the 
lowest order observable which tests the whole function $h(x)$ for $0 <
x< L$. It appears as a fundamental quantity in disordered systems.

The aim of this Letter is to compute the width distribution (WD)
$\Phi(z)$ for elastic interfaces driven in random media, exactly at
the depinning threshold, numerically and from field theory. As in the
linear problems treated earlier, we confirm the existence of universal
properties in various dimensions $d$ and with several functional forms of
the elasticity.  The surprising finding is that in all cases $P(w^2)$
(i.e.\ its shape) is {\em extremely well} approximated by a simple
generalized Gaussian approximation (GA), without any  fit parameter,
and depends only on $\zeta$, which is determined independently.
This suggests that the complicated morphology of interfaces (cracks,
domain walls, etc.) may be rendered by a simple ansatz of independent modes
with a characteristic decay. This may have important consequences for the
analysis of numerical and experimental data.  Our numerical results are
then understood within a functional renormalization group calculation,
detailed in a companion paper \cite{technical}.

We consider the zero temperature equation of motion of an interface
given by
\begin{equation} 
\partial_t h(\VEC{x},t) = - \frac{\partial E }{\partial h(\VEC{x},t) }
= f + \eta(\VEC{x},h (\VEC{x},t)) - \frac{\partial
E_{\text{el}}}{\partial h(\VEC{x},t)} \ ,\label{e:continuum_motion}
\end{equation}
where the functional $E (\left\{h,\VEC{x} \right\})$ represents the
total energy incorporating potential energy due to the driving force
$f$, the  short-range correlated  disorder force
$\eta(\VEC{x},h)$, as well as its internal convex elastic energy
$E_{\text{el}}$.  Equation~(\ref{e:continuum_motion}) is non-linear,
and has not been solved exactly. We are interested in the depinning
limit ($f=f_c$) where the velocity of the elastic manifold goes to
zero. We use periodic boundary conditions, and recall that the WD
$\Phi(z)$, although independent of small scale details, does depend on
the boundary condition at large scale.

For our numerical study we use a very efficient algorithm
\cite{Krauth.zeta.01,Rosso.MonteCarlo.02,Rosso.contact.02} which
directly determines critical forces $f_c$ as well as the critical
interface $h_c(\x)$ for a wide range of models. In particular we
calculate the WD for interfaces of dimensions $d=1$ and $d=2$, where
the elastic energy has the harmonic form $E_{\mathrm{el}} (\{h,\x
\})\sim(\nabla h)^{2}$.  We have also tested the universality of
$\Phi(z)$ in $d=1$, by means of a directed polymer model with an
anharmonic quartic elasticity, and for a lattice model with hard local
constraint, which have the same $\zeta=0.63$ \cite{Krauth.zeta.01}. As
expected, $\Phi(z)$ is always size independent and the WD associated
to non-harmonic models can be distinguished from the one resulting
from an harmonic elasticity.  The harmonic models, in fact, have an
exponent $\zeta=1.2$, and thus belong to a different universality
class.

For our field theory calculation we use the functional renormalization
group method (FRG) originally developped to one loop to describe
the model with harmonic elasticity and correct the predictions of
dimensional reduction \cite{Nattermann.pinning.97,Narayan.Pinning.93}.
Recently a renormalized field theory was constructed to 2-loop order
\cite{Wiese.2loop.01} which overcomes the deficiencies of the 1-loop
analysis; notably to distinguish between statics and driven dynamics, and
to account for the large values of the roughness exponent $\zeta$ measured
e.g.\ in \cite{Rosso.MonteCarlo.02,Rosso.contact.02,Rosso.manifolds.02}
as compared to an earlier conjecture \cite{Narayan.Pinning.93} $\zeta =
(4-d)/3$. We find that the FRG both suggests the GA as a lowest order
approximation in $\epsilon=4-d$ and allows to define and compute universal
ratios which probe high cumulants of $P(w^2)$ and deviations from the GA,
and are thus more sensitive to details of the universality class. The
simplest of them is the generalized kurtosis
\begin{equation}\label{kurtosis}
R = \frac{\int_{x,y}\sample{\left ( u(x)^{2} u(y)^{2}
\right)}^{\mathrm{c}}}{2\int_{x,y}\left (\sample{ u(x)u(y)} \right)^{2}}\ ,
\end{equation}
where the subscript $\mathrm{c}$ indicates the connected expectation value.  $R$
is found to be small but non-zero. This directly proves that the correct
description of interfaces must go beyond the independent-mode picture.

To introduce the Gaussian approximation in the most elementary way, we
first recall \cite{Racz.Random.94} the simple periodic random
walk of size $L$, with a Fourier expansion
\begin{equation}
u(x)= \sum_{n=1}^{\infty} a_n \cos\left(\frac{2 \pi n}{L}x \right) 
 + b_n \sin\left( \frac{2 \pi n}{L}x \right)\ .
\label{e:Fourier}
\end{equation} 
The standard Gaussian probability measure associated with $u$, i.e.\
${\cal P}[u] \propto \exp[-\frac{1}{2}\int_0^L \rmd x (\frac{\partial
u(x)}{\partial x})^2]$ gives
\begin{equation}\label{e:Gaussian}
{\cal P}[u] 
\propto \exp\left[- \sum_{n=1}^{\infty} \frac{( \pi n)^2}{L}
(a_n^2+b_n^2) \right]\ .  
\end{equation}
The probability distribution $P(w^2)$
\begin{equation}
P(w^2)=\int {\cal D}[u] \delta(w^2 - \mean{u^2})\, {\cal P}[u]
\label{e:PDF}
\end{equation}
is obtained from the  generating function of its moments
\begin{equation}\label{e:generatrice}
W(\lambda) =  \int_{0}^{\infty} \rmd w^2  P(w^2) \rme^{-\lambda w^2}.
\end{equation}
Writing Eq.~(\ref{e:generatrice}) as an integral over $a_{n}$ and
$b_{n}$, we obtain
\begin{eqnarray}\label{e:generatricewalk}
&& Z(\lambda) = \prod_{n=1}^\infty \int \rmd a_{n}\, \rmd b_{n}\, \rme^{-\frac{(\pi
n)^{2}}{L}
(a_{n}^{2}+b_{n}^{2})} \rme^{-\frac{\lambda}{2}
(a_n^{2}+b_n^{2})}\nonumber\\ 
&& W(\lambda) = \frac{Z(\lambda)}{Z(0)} = \prod_{n=1}^\infty \left( 1 + \frac{\lambda}2 \frac{L}{(\pi n)^{2}} \right)^{\!-1}.
\end{eqnarray}
For the random walk Eq.~(\ref{e:Gaussian}), 
$P(w^2)$ can be obtained exactly by inverse Laplace transform of
Eq.~(\ref{e:generatricewalk}):
\begin{equation} 
P(w^2)=
\frac{4 \pi^{2}}L \sum_{n>0} n^{2} (-1)^{n+1} \,\rme^{-2 {w^{2}} 
(\pi n)^{2}/L}.
\label{e:PDFrw}
\end{equation} 
Using $\sample{w^2}=-\frac{\rmd W}{\rmd \lambda}|_{\lambda = 0}=\frac{L}{12}$,
Eq.~(\ref{e:PDFrw}) can be written in a scaling form
\begin{eqnarray}\label{e:scaling}
 \Phi (z)&=&\sample{w^2}   P({w^2})\ , \qquad z =
w^2/\sample{w^2}\nonumber  \\
&=& \frac{\pi^{2}}{3} \sum_{n>0} n^{2} (-1)^{n+1}\rme^{-
\frac{\pi^{2}}{6}z n^{2}}.
\end{eqnarray}
The size dependence thus appears only through the average width
$\sample{w^2}$.
We can generalize Eq.~(\ref{e:Gaussian}), where each mode $a_n, b_n$
has a  weight $\propto n^2$, to an arbitrary function of
independent Fourier modes
\begin{equation} 
\label{e:GaussianGenN}
{\cal P}_{\g}[u] \propto  \exp\left[ -\frac{L}{4} \sum_{n>0} (a_n^2 +b_n^2)\,
G^{-1}\!\left(\frac{2 \pi n}{L}\right)  \right]  \ .
\end{equation}
which, in real space, corresponds to 
\begin{equation} 
\label{e:GaussianGen} {\cal P}_{\g}[u] \propto \exp\left[ -\frac{1}{2}
\int_0^L \int_0^L \rmd x\, \rmd y\, u(x) G^{-1}_{xy}\, u(y) \right] \ .
\end{equation}
The function $G_{xy} = \sample{u(x) u(y)}$ is the exact disorder-averaged
2-point function and can be computed from numerical data.  This
allows to obtain ${\cal P}_{\g}$ even for a finite system.  In the
thermodynamic limit, ${\cal P}_{\g}[u]$ is obtained from the behavior
of $G_{xy}=G_{x-y}$ for large  $|x-y|$ (small $q$),  where $ G(q) \sim
C/q^{d + 2 \zeta}$.  This means that a single observable,  $\zeta$,
fixes ${\cal P}_{\g}[u]$ on large scales.

We again determine the generating function for the moments, but for 
arbitrary $\zeta$ and $d$:
\begin{equation}
W(\lambda) = \prod_{q \neq 0} \left( 1 + 2 \tilde \lambda G(q)
\right)^{-1/2} \ ,  \label{e:final}
\end{equation}
where $\tilde \lambda= \lambda/L$, $q=2 \pi n/L$, $n \in \mathbb{Z}^d$.  Due to
the symmetry $q \leftrightarrow -q$, no fractional power appears in
Eq.~(\ref{e:final}), as in Eq.~(\ref{e:generatricewalk}), where the exponent 
$-1$ stems from the double sum over the $a_n$ and $b_n$.  
An explicit sum over poles  allows to obtain
$\Phi_{\g}(z)$ for all  $\zeta$ and $d$  with excellent precision.
All GA interfaces $\{u(x)\}$ can be directly  sampled by Monte Carlo methods.
For details, including the extension to open boundary conditions, see
\cite{Antal.1overf.01,RossoKrauth03}

\begin{figure}[t]
\centerline{\Fig{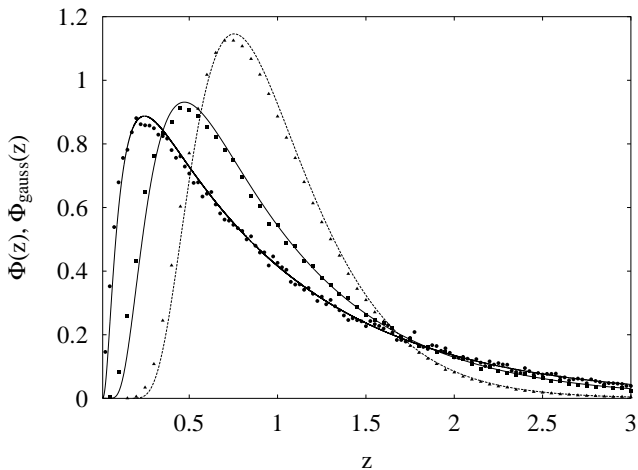}}
\caption{Scaling function $\Phi(z)$ and $\Phi_{\g}(z)$ for:
($1+1$)--d harmonic ($L=256$, $2 \times 10^5$ samples, $\z=1.25$)---left; 
($1+1$)--d an-harmonic ($L=256$, $2 \times 10^5$ samples, $\z=0.63$)---middle; 
($2+1$)--d harmonic ($L=32$, $10^5$ samples, $\z=0.75$)---right.
The scatter in the numerical data is mostly due to binning. 
Notice that, for $d=1$, the typical value of $z$ is much
smaller than its average $\sample{z}=1$.} 
\label{f:Dhm}
\end{figure}

In Fig.~\ref{f:Dhm} we compare, for different models, the exact scaling
function $\Phi(z)$ to $\Phi_{\g} (z)$, using $G(n)= C/n^{d+2 \z}$. The
roughness exponent was previously obtained  using both field theory
\cite{Wiese.2loop.01} and numerical methods \cite{Rosso.MonteCarlo.02}.
The agreement between $\Phi$ and $\Phi_{\g}$ is clearly spectacular.
The scatter of the data, visible in Fig.~\ref{f:Dhm}, is mostly due to
the finite width of histogram bins.

Tiny---yet significant---differences between $\Phi$ and $\Phi_{\g}$
are best resolved in the integrated probability distributions,
which need no discretization. The difference between the integrated
distributions is
\begin{equation} 
\Delta H(z) =  \int_0^{z} \rmd t (\Phi_{\g}(t) - 
\Phi(t)) , 
\label{e:Diff} 
\end{equation}
where $H(z)=\int_0^{z} \rmd t  \Phi(t)$  is the fraction of samples
with a renormalized width below $z$. 
\begin{figure}[t]
\centerline{\Fig{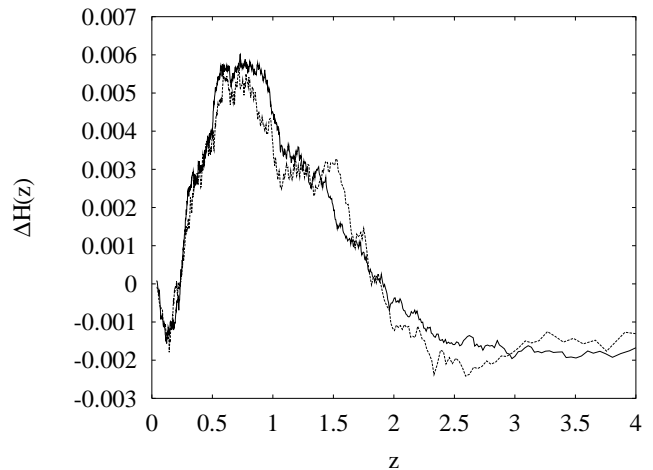}}
\caption{Difference between the integrated distribution functions of $\Phi$ and
$\Phi_{\g}$ (with $\z=1.25$)
obtained from $2 \times 10^5$ independent interfaces at
$L=256$ (continuum line) and $L=64$ (dashed line), in the $d=1$ harmonic
model. 
}
\label{f:Diff}
\end{figure}
In Fig.~\ref{f:Diff}, we show $\Delta H(z)$ obtained from $N=2 \,
10^5$ independent samples.  Statistical fluctuations in this quantity
are of order $1/\sqrt{N}$ and the signal-to-noise ratio would be
smaller than $1$ if $N$ was an order of magnitude smaller.  The
absence of systematic finite-size effects shows that the asymptotic
regime of large interfaces has been reached and thus to conclude that
the exact distribution for large systems is not
Gaussian.

We now  discuss the field theoretical calculation.  To lowest
order in perturbation theory, we show that the generalized Gaussian
approximation appears naturally. This is instructive since it identifies
the diagrams which are obtained by assuming the theory to be Gaussian,
albeit non-trivial, since it involves a non-trivial roughness exponent
$\zeta$.  Using dynamical field theoretic methods \cite{Wiese.2loop.01},
one starts again from the Laplace transform $W(\lambda)$ and expands in
powers of the correlator of the pinning force $\Delta(u)$.  To lowest
order one finds \cite{technical} that $\log W(\lambda)$ is the sum of all
connected 1-loop diagrams.  The loop with $N$ disorder vertices and $N$
insertions of $w^2$ is
\begin{equation}
\frac1{2 N} \sum_q \left(\frac{- 2 \tilde \lambda \Delta(0)}{(q^2)^2} \right)^N , 
\label{one}
\end{equation}
where the sums over $q$ thus run over a $d$-dimensional lattice with
spacing $\frac{2\pi}{L}$, and the 0-mode is excluded, as appropriate
for periodic BC.  Resumming (\ref{one}) over $q$ would give
Eq.~(\ref{e:final}) with $G(q) \sim 1/q^4$, i.e.  the dimensional
reduction (Larkin) result. In fact, the FRG tells us that the
calculation should be performed with the running disorder $\Delta(0)
\to \Delta_l(0) = \rme^{(\epsilon - 2 \zeta ) l} \tilde{\Delta}^*(0)$
where $\tilde{\Delta}^*(0)$ is the (non-universal) value of the fixed
point \cite{Wiese.2loop.01}. For the present
case of periodic boundary conditions and momentum infrared cutoff, 
one can replace $l \to
\log(1/q)$, and finally obtains Eq.~(\ref{e:final}) with $G(q) = C/q^{d
+ 2 \zeta}$.  This calculation is valid to dominant order in
$\epsilon=4-d$, i.e.\ near $d=4$. If the same class of diagrams are
resummed in any $d$ it  leads to the GA, as we now
illustrate considering e.g.\ the second connected cumulant of the
WD. This cumulant is {\em not connected} w.r.t.\ $h$, and thus there
is an exact relation:
\begin{equation}
\sample{(w^2)^2}^c  =  \sample{ (w^2)^2 } - \sample{(w^2)}^2 =
 2\left( 1 + R \right)\int_{x , y}  G_{xy}^2.
\label{e:conn}
\end{equation}
The first term results from Wick's theorem and would be the full
result if the measure were Gaussian.  Analogous formulae exist for
higher cumulants, and if the measure of $h$ is purely Gaussian can be
resummed into Eq.~(\ref{e:final}). Even though the GA is not exact,
the deviations, given by the last term in (\ref{e:conn}) are expected
to be small; indeed they are of order $\epsilon^4$. Thus the GA is
already exact to the {\it two lowest leading orders} $\epsilon^2$ and
$\epsilon^3$, which explains why it is so accurate even in low
dimension.

The calculation of the deviations using the field theory is delicate
\cite{technical}.  The kurtosis $R$ in Eq.~(\ref{kurtosis}) which
characterizes the importance of non-Gaussian effects is found to be $R =
- 0.13 \epsilon^2$ to lowest order for small $\epsilon=4-d$. It is easy
to see that this strongly overestimates $R$ in low dimensions. Another
method is to work in fixed dimension and to truncate to one loop, yielding
$R=- 0.036$ ($d=3$), $R=- 0.048$ ($d=2$), $R= - 0.01$ ($d=1$), which in view 
of the numerical results below seems to underestimate $R$. The small
values obtained in low dimensions arise from kinematic constraints in
the diagrams, presumably a genuine effect indicating large corrections
from higher orders in $\epsilon$ to the above $O(\epsilon^2)$ even in
$d=3$. Note that the {\it sign} of the result indicates a distribution
more peaked than a Gaussian and is in agreement with the only other known
exact result \cite{cecile} for the (random field) statics in $d=0$, $R=
- 0.080865..$.

We have computed from Eq.~(\ref{kurtosis}) the generalized
kurtosis function, in a model-independent way. 
We have checked on the $1-d$
harmonic model that, as $\Phi(z)$, $R$ is not affected by finite size
effects and -- using $10^6$ samples -- we find:
$R=-0.054 \pm 0.002$ ($1-d$ harmonic $L=256$);
$R=-0.067 \pm 0.002$ ($1-d$ anharmonic $L=64$);
$R=-0.053 \pm 0.002$ ($2-d$ harmonic $L=32$).
As proven by FRG calculations, $R$ is small but definitely different from
zero. Direct information on the non-Gaussian effects can also be obtained from
the Fourier transforms of the interfaces $u(x)$ in Eq.~(\ref{e:Fourier}).
In Fourier space, for $d=1$, the expression of $R$ is: 
\begin{equation}
R=  \frac{\sum_{n_1 , n_2} \sample{ (a_{n_1}^2 +b_{n_1}^2) (a_{n_2}^2 +b_{n_2}^2)   }^c}
{2  \sum_{n} \sample{ (a_{n}^2 +b_{n}^2)}^2}.
\label{e:}
\end{equation}
We remark that $R$ detects correlations in the disorder-averaged 
fourth moments $\sample{|u(q_1)|^2 |u(q_2)|^2}$,
which cannot be expressed simply through the second moments
$\sample{|u(q)|^2}$.

To summarize, we have computed both numerically and within field theory
the width distribution of critical configurations at depinning, with
consistent results.  The shapes of the distributions are strongly
dominated by the value of $\zeta$.  On the other hand, it will
be difficult to distinguish different universality classes from
the forms of $\Phi(z)$, if their roughness exponents are similar.
Other universal quantities, such as $R$ defined here, directly
involve the  non-Gaussian part of the distribution. Their precise
determination however requires more work, both numerically and within
field theory. Also, since the WD is so tied up to $\z$, finite size
effects in both quantities are connected.  Finite-size effects
will need to be  well understood in order to resolve open issues
\cite{Buldyrev.anisotropic.93,Rosso.manifolds.02,LeDoussalWiese2002a}
concerning $d_{\mathrm{uc}}$ for the anisotropic depinning class. It
would be interesting to carry similar calculations on other pure, and
disordered models, in particular for the equivalent static system.

We thank Z.~R\'acz for stimulating discussions. K.J.W.~is supported by
Deutsche Forschungsgemeinschft under grant Wi 1932/1-1.

\end{document}